\documentclass[onecolumn]{IEEEtran} 

\usepackage{graphicx}
\usepackage{comment}
\usepackage{here}
\usepackage{wrapfig}
\usepackage[export]{adjustbox}
\usepackage{amsmath}
\usepackage{algorithm,algpseudocode}
\usepackage{caption}

\usepackage{multicol}
\usepackage{url}

\title{Assembling a Pipeline for 3D Face Interpolation}
\author{Yusuke Niiro $<$yniiro@ucmerced.edu$>$ \\ Marcelo Kallmann $<$mkallmann@ucmerced.edu$>$ \\
University of California, Merced}

\begin{document}

\maketitle

\begin{abstract}
This paper describes a pipeline built with open source tools for interpolating 3D facial expressions taken from images.
The presented approach allows anyone to create 3D face animations from 2 input photos: one from the start face expression, and the other from the final face expression.
Given the input photos, corresponding 3D face models are constructed and texture-mapped using the photos as textures aligned with facial features.
Animations are then generated by morphing the models by interpolation of the geometries and textures of the models.
This work was performed as a MS project at the University of California, Merced.
\end{abstract}

\begin{multicols*}{2}
	
\section{Introduction}

It is common to observe many realistic 3D animated characters being used in video games and movies. 
Behind the scenes, large data sets of images are often used in order to achieve  realistic appearances.
Designers are also often responsible for editing the characters in order to improve the final results.
While significant advances have been achieved in facial animation, it is still difficult to make human-like 3D faces that always look realistic.
People are very good at recognizing the differences between real humans and digital humans.

3D character animation technologies are also starting to be used in a variety of new innovative products. 
For example, the newest iPhone (smartphone by Apple) has an application that can take pictures or videos of people and place 3D models covering their faces and moving in coordination with the real faces.
Such an application illustrates the growing need for simple and efficient approaches for achieving realistic face animation.
This paper describes a pipeline built with open source tools for achieving animations by interpolation of 3D facial expressions taken from pictures.


\section{Related Work}

A significant amount of previous work has relied on datasets of human faces in order to build face models.
For example, the approach of Blanz and Vetter~\cite{3dmm} applied pattern classification on their dataset of human faces in order to reconstruct a 3D face model from a single 2D face image.
Booth et al.~\cite{largescale} introduced the Large Scale Facial Model (LSFM), which is able to automatically construct 3D models of a variety of human faces from a data set of 9,633 distinct facial identities. 
Tran et al.~\cite{nonlinear} also proposed a framework to construct 3D models from a large set of face images, but without requiring to collect 3D face scan data. 
Huber et al. \cite{framework} presented the Surrey Facial Model, a multi-resolution model that can build a facial model in different resolution levels.

Facial animation can be accomplished by various approaches.
Chuang and Bregler \cite{blendshape} describe an approach for facial animation that is based on motion capture data and interpolation of blend shapes. 
Noh and Neumann \cite{cloning} proposed the work Expression Cloning, an approach that makes a facial movement by transferring vertex motion data from one source to another. 
Lee et al. \cite{muscle} introduced an approach to generate facial expressions based on muscle information from real face data.

There are also available products that allow the creation and manipulation of 3D facial models.
For example, Face Poser is a system presented by Lau et al.~\cite{faceposer} and Poser~\cite{poser} is a software  system that facilitates modeling and editing 3D faces with a comprehensive graphical user interface.

In this work a solution based on available open source tools is presented.
3D face models are created from photos based on the Surrey Facial Model~\cite{framework}.
In order to achieve a simple approach for face animation, this work focuses on animating the interpolation between a pair of input facial expressions.





\subsection{Implementation Details}

This project was implemented in C++ with Visual Studio 2017 under Windows 10.
The following C++ libraries were used:

\begin{itemize}

\item dlib library \cite{dlib}: 
This library is used to process the input face images and to place feature landmarks on the images.
The landmarks are based on the ibug~\cite{ibug} facial points (Fig. \ref{one}), which represent 68 feature landmarks for human faces. These features are later used to generate the 3D face models and to morph 2D images (using the eos library~\cite{eos}). 
Figures \ref{two} and \ref{three} illustrate example images including their respective facial landmarks.

\begin{figure}[H]
	\begin{center}
		\includegraphics[clip, width=5.0cm]{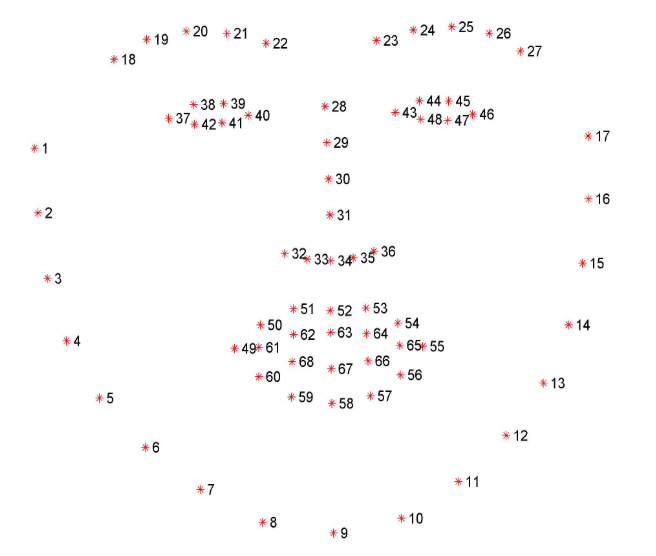}
		\caption{Facial landmarks.}
		\label{one}
	\end{center}
\end{figure}

\begin{figure}[H]
	\begin{center}
		\includegraphics[clip, width=8.0cm]{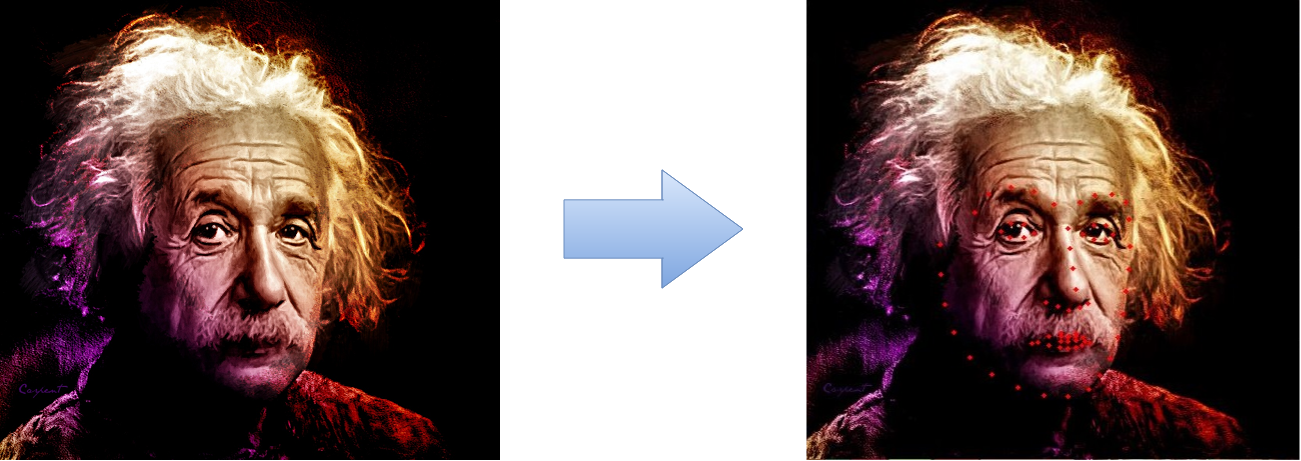}
		\caption{Facial landmarks on Albert Einstein's face.}
		\label{two}
	\end{center}
\end{figure}

\begin{figure}[H]
	\begin{center}
		\includegraphics[clip, width=8.0cm]{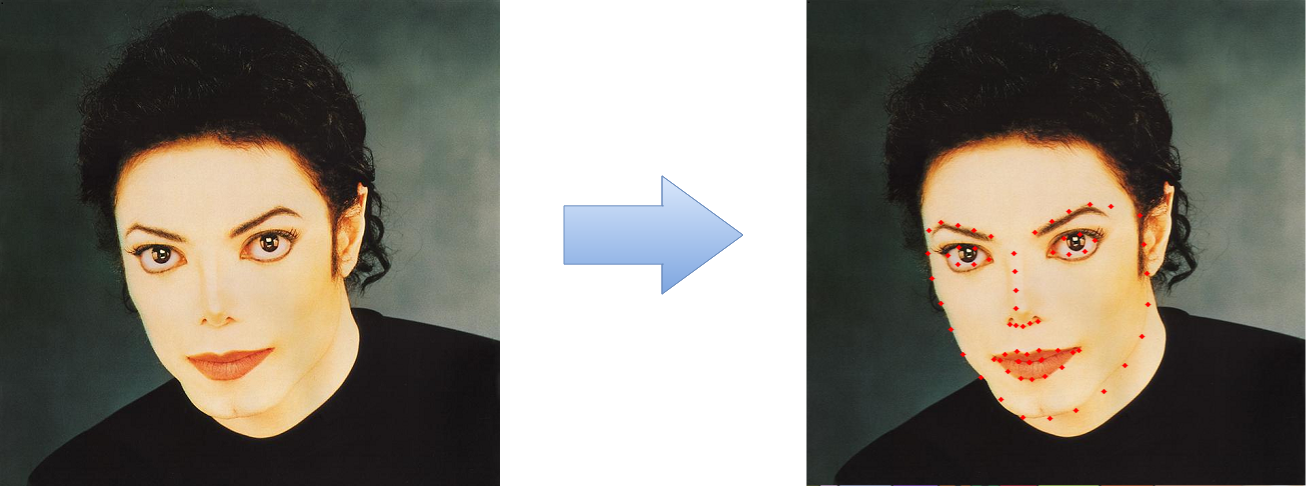}
		\caption{Facial landmarks on Michael Jackson's face.}
		\label{three}
	\end{center}
\end{figure}

\item eos library \cite{eos}: 
This library is used to create 3D face models from input photos of human faces.
The obtained models follow the Surrey Facial Model, which is based on principal component analysis applied to a large data set of face models. 
These models can be created in several resolutions. 
In this project we experimented with face models built with 3448, 16759, and 29587 vertices.
This library depends on the OpenCV, Eigen, and Boost libraries.
Figures~\ref{four} and \ref{five} illustrate example models which were generated from their corresponding input photos.

\begin{figure}[H]
	\begin{center}
		\begin{tabular}{c}		
			\begin{minipage}{0.5\hsize}
				\begin{center}
					\includegraphics[clip, width=4.0cm]{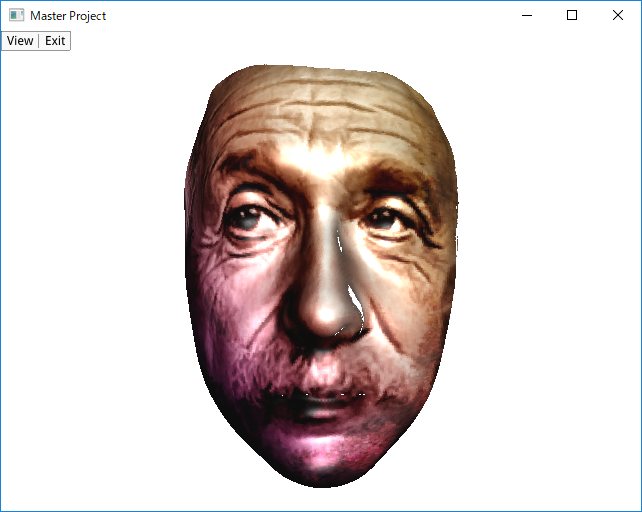}
					\hspace{1.6cm} [1] Front view
				\end{center}
			\end{minipage}
			\begin{minipage}{0.5\hsize}
				\begin{center}
					\includegraphics[clip, width=4.0cm]{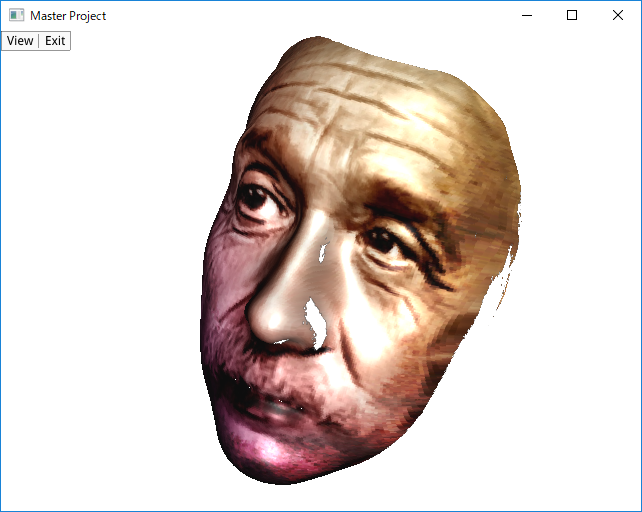}
					\hspace{1.6cm} [2] Side view
				\end{center}
			\end{minipage}
		\end{tabular}
		\caption{3D Face Model of Albert Einstein.}
		\label{four}
	\end{center}
\end{figure}

\begin{figure}[H]
	\begin{center}
		\begin{tabular}{c}		
			\begin{minipage}{0.5\hsize}
				\begin{center}
					\includegraphics[clip, width=3.8cm]{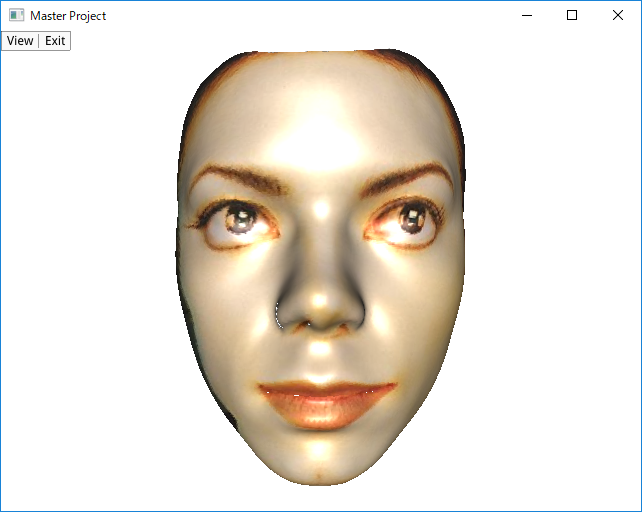}
					\hspace{1.6cm} [1] Front view
				\end{center}
			\end{minipage}
			\begin{minipage}{0.5\hsize}
				\begin{center}
					\includegraphics[clip, width=3.8cm]{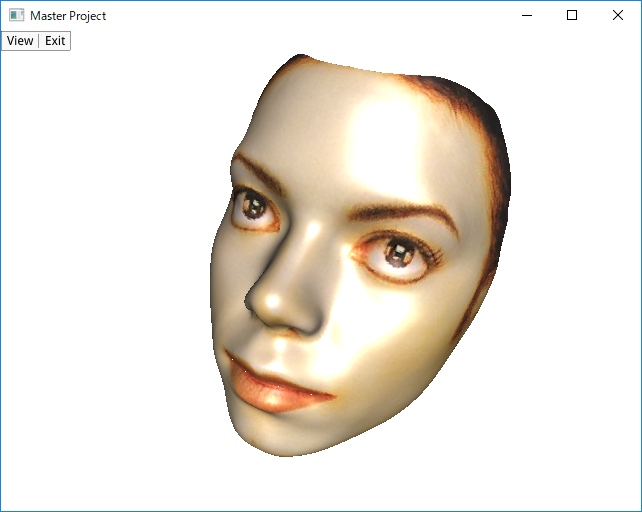}
					\hspace{1.6cm} [2] Side view
				\end{center}
			\end{minipage}
		\end{tabular}
		\caption{3D Face Model of Michael Jackson.}
		\label{five}
	\end{center}
\end{figure}

\item OpenCV library \cite{opencv}:
This library was used to support image operations.
In particular, this library was used to triangulate the facial landmarks in order to interpolate image attributes based on triangle coordinates, such that the interpolated image information preserves the facial features, as later explained in Section~\ref{sec:triang}.
Figure~\ref{six} illustrates the obtained triangulations in example photos and Figure~\ref{seven} illustrates the obtained interpolation result preserving the facial features.

\begin{figure}[H]
	\begin{center}
		\begin{tabular}{c}		
			\begin{minipage}{0.5\hsize}
				\begin{center}
					\includegraphics[clip, width=3.8cm]{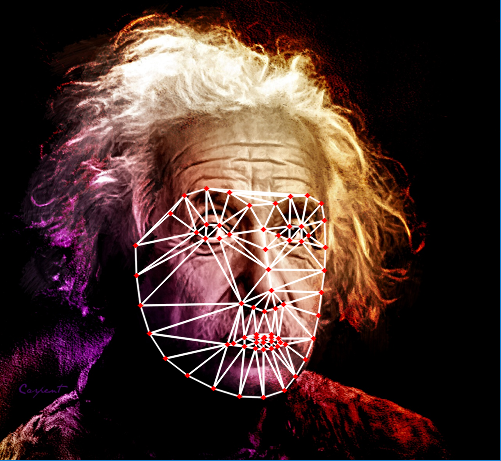}
					\hspace{2.0cm} [1] Albert Einstein
				\end{center}
			\end{minipage}
			\begin{minipage}{0.5\hsize}
				\begin{center}
					\includegraphics[clip, width=3.8cm]{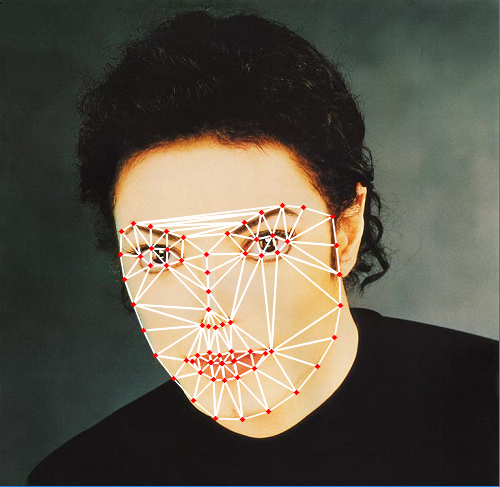}
					\hspace{2.0cm} [2] Michael Jackson
				\end{center}
			\end{minipage}
		\end{tabular}
		\caption{Triangulation on example photos.}
		\label{six}
	\end{center}
\end{figure}

\begin{figure}[H]
	\begin{center}
		\includegraphics[clip, width=1\columnwidth]{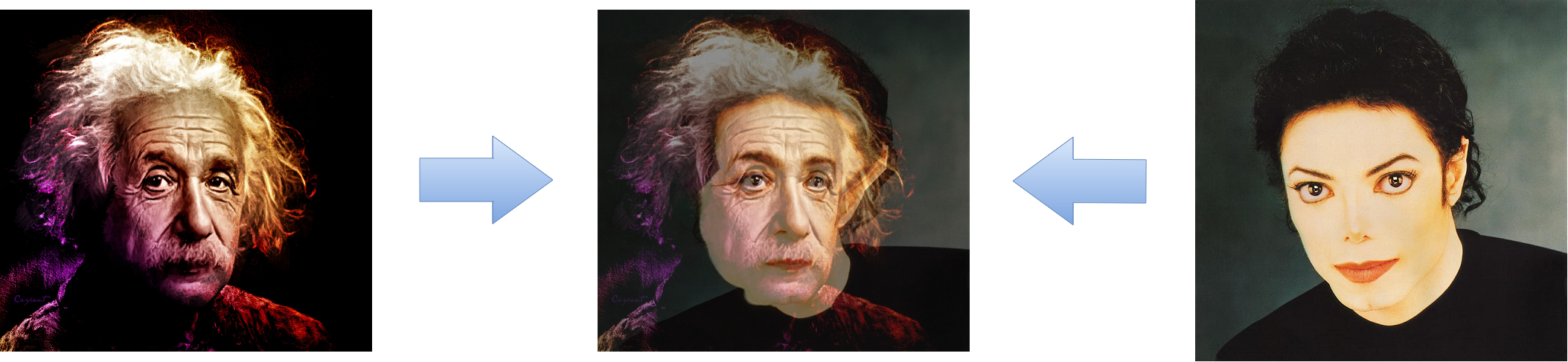}
		\caption{Morphed image of 2 example photos. }
		\label{seven}
	\end{center}
\end{figure}

\item Eigen library \cite{eigen}:
This library provides functions for linear algebra, matrix operations, geometrical transformations, numerical solvers and related algorithms.

\item Boost library \cite{boost}: 
This library provides various functions to complement the standard C++ libraries.

\item Standalone Interactive Graphics (SIG) library \cite{sig}:
This library provides a C++ scene graph framework for the development of applications with interactive 3D graphics.
The main animation visualizer was built with this library, which handled the graphical user interface and scene graph operations including computing the performed interpolations handling images and 3D objects.

\end{itemize}

\section{Overall Method}

\begin{figure}[H]
	\begin{center}
		\includegraphics[scale = 0.14]{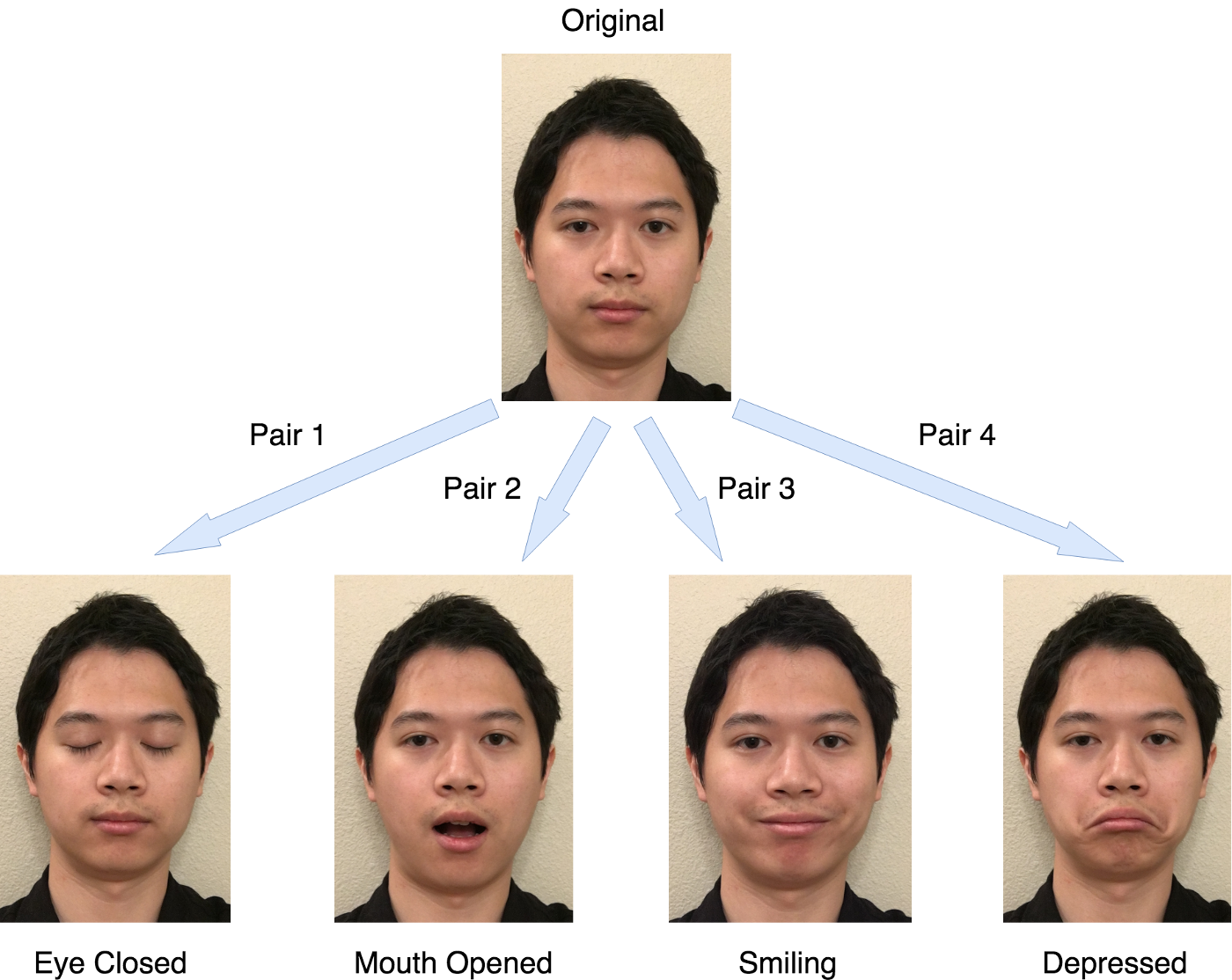}
		\caption{Four pairs of combination used in this project (Pair 1: original photo and eye closed photo, Pair 2: original photo and mouth opened photo, Pair 3: original photo and smiling (mouth up) photo, Pair 4: original photo and depressed (mouth down) photo). \label{fig:eight}}
	\end{center}
\end{figure}

Our overall method consists of three main steps.
First, 3D face models from pairs of input photos are created. 
In this work we have used 4 pairs in order to obtain 4 animations between different facial expressions.
These pairs are presented in Figure~\ref{fig:eight}.

Second, for each pair, the corresponding texture images are morphed in order to obtain a texture image for a given interpolation factor $t \in [0,1]$, where $t$ allows to obtain results from the initial model ($t=0$) to the goal model ($t=1$). 

Finally, the coordinates of the 3D models are interpolated according to parameter $t$ and associated with the intepolated texture image computed in the previous step.
These steps are detailed in the next subsections.


\subsection{Generation of 3D Face Models}

\begin{figure}[H]
	\begin{center}
		\begin{tabular}{c}		
			\begin{minipage}{0.5\hsize}
				\begin{center}
					\includegraphics[clip, width=3cm]{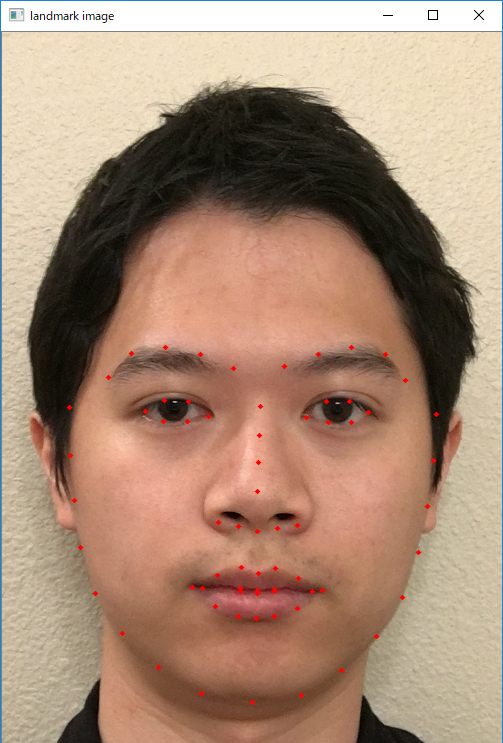}
					\hspace{3.0cm} [1] Original photo
				\end{center}
			\end{minipage}
			\begin{minipage}{0.5\hsize}
				\begin{center}
					\includegraphics[clip, width=3cm]{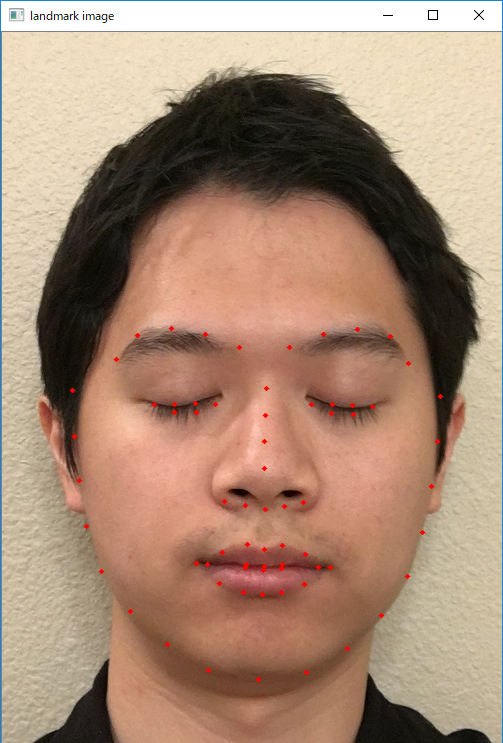}
					\hspace{3.0cm} [2] Eye closed photo
				\end{center}
			\end{minipage}
		\end{tabular}
		\caption{Facial landmarks placed on the input photos.}
		\label{ten}
	\end{center}
\end{figure}

The input 3D face models are generated from 2
photos of the same person but with different facial expressions. 
One of the photos is used as the initial face and the other photo is used as the goal face.
At this point the coordinates of the facial landmarks are computed for each photo by using the dlib library. 
Figure~\ref{ten} illustrates one pair of photos with the facial landmarks placed on them.



\begin{figure}[H]
	\begin{center}
		\begin{tabular}{c}		
			\begin{minipage}{0.5\hsize}
				\begin{center}
					\includegraphics[clip, width=3.8cm]{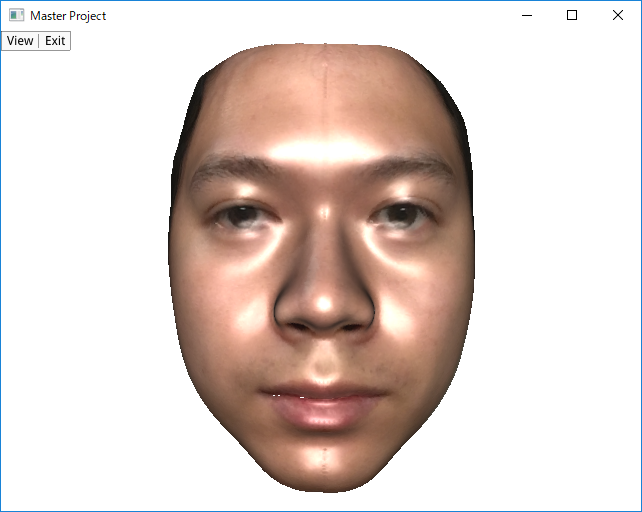}
					\hspace{3.8cm} [1] Front view
				\end{center}
			\end{minipage}
			\begin{minipage}{0.5\hsize}
				\begin{center}
					\includegraphics[clip, width=3.8cm]{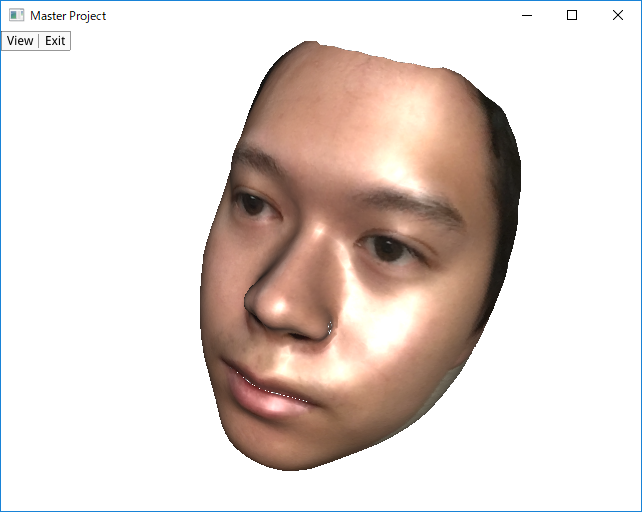}
					\hspace{3.8cm} [2] Side view
				\end{center}
			\end{minipage}
		\end{tabular}
		\caption{3D Face Model based on original photo.}
		\label{eleven}
	\end{center}
\end{figure}

\begin{figure}[H]
	\begin{center}
		\begin{tabular}{c}		
			\begin{minipage}{0.5\hsize}
				\begin{center}
					\includegraphics[clip, width=3.8cm]{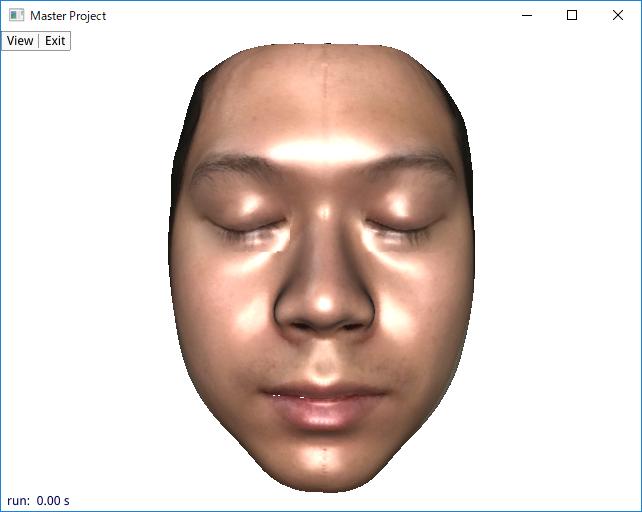}
					\hspace{3.8cm} [1] Front view
				\end{center}
			\end{minipage}
			\begin{minipage}{0.5\hsize}
				\begin{center}
					\includegraphics[clip, width=3.8cm]{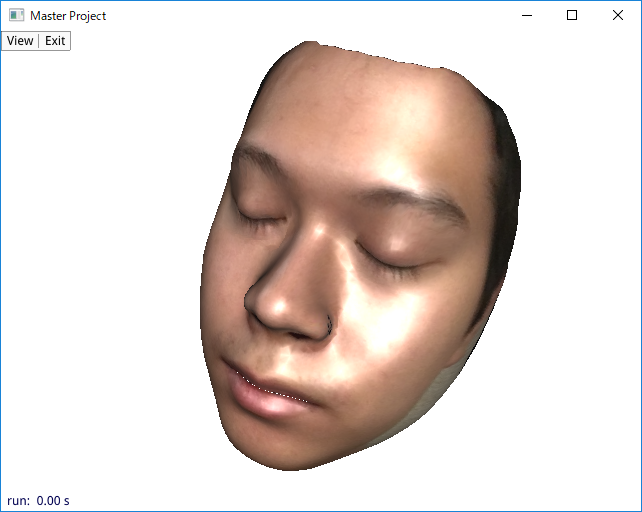}
					\hspace{3.8cm} [2] Side view
				\end{center}
			\end{minipage}
		\end{tabular}
		\caption{3D Face Model based on eye closed photo.}
		\label{twelve}
	\end{center}
\end{figure}

The 3D facial models are then generated by using the original photos together with the facial landmark information configured in the previous step by using the eos library.
Figures~\ref{eleven} and \ref{twelve} illustrate the 3D face models generated.

\subsection{Morphed Texture Images}\label{sec:triang}

\begin{figure}[H]
		\begin{center}
			\begin{tabular}{c}		
				\begin{minipage}{0.5\hsize}
					\begin{center}
						\includegraphics[clip, width=3.0cm]{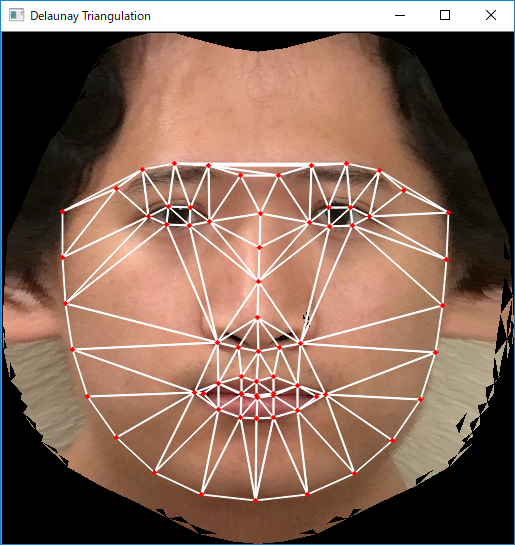}
						\hspace{3.8cm} [1] Texture image of\\ original 3D model
					\end{center}
				\end{minipage}
				\begin{minipage}{0.5\hsize}
					\begin{center}
						\includegraphics[clip, width=3.0cm]{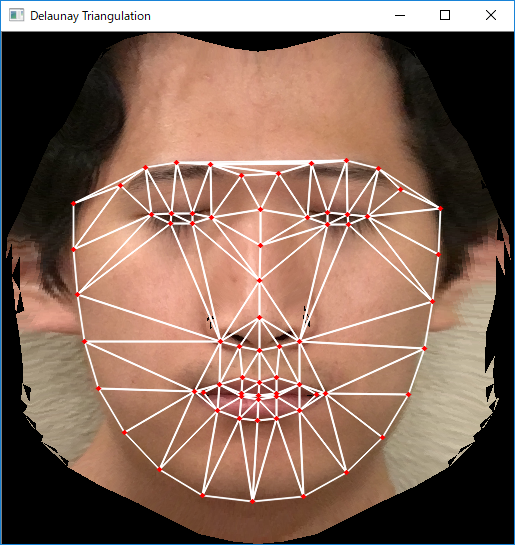}
						\hspace{3.8cm} [2] Texture image of\\ eye closed 3D model
					\end{center}
				\end{minipage}
			\end{tabular}
			\caption{Triangulation on texture images.}
			\label{fourteen}
		\end{center}
	\end{figure}

\begin{figure}[H]
		\begin{center}
			\includegraphics[clip, width=8.0cm]{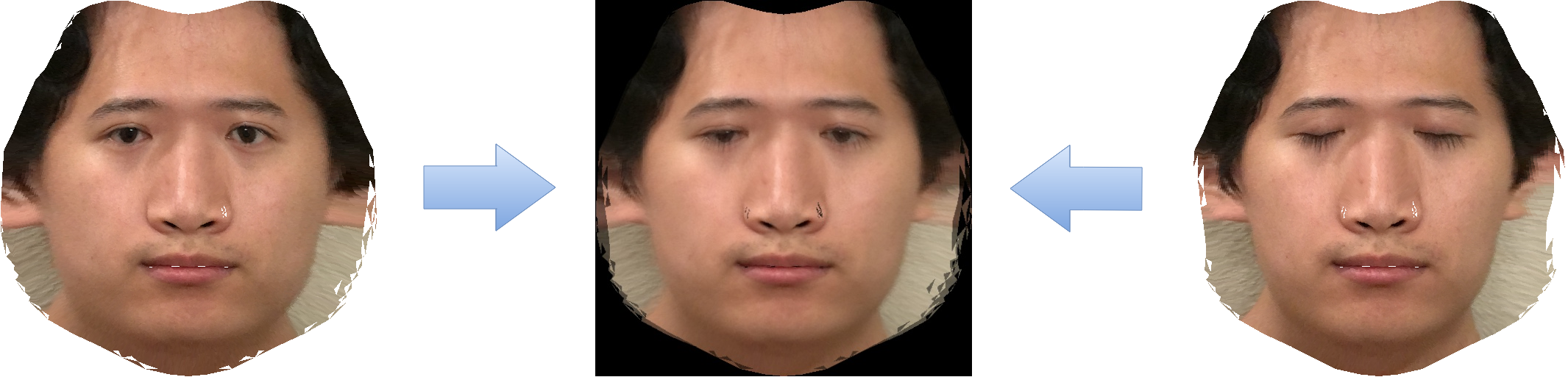}
			\caption{Morphed image of original and eye closed images.}
			\label{fifteen}
		\end{center}
	\end{figure}

To morph the texture images, 
the 
coordinates of the facial landmarks are configured for each texture image by using the dlib library. 
The landmarks are then triangulated using a Delaunay triangulation (Fig.~\ref{fourteen}),
such that the coordinates of the image texture are interpolated using barycentric coordinates according to the position of each image pixel in its containing triangle. 
The colors of the image texture are also interpolated in the same way.
In this way the facial features are preserved during interpolation.
Figure~\ref{fifteen} illustrates morphed image of original and eye closed texture images of 3D face models.


\subsection{Interpolation of Mesh Coordinates}



\begin{figure}[H]
	\begin{center}
		\begin{tabular}{c}		
			\begin{minipage}{0.5\hsize}
				\begin{center}
					\includegraphics[clip, width=3.8cm]{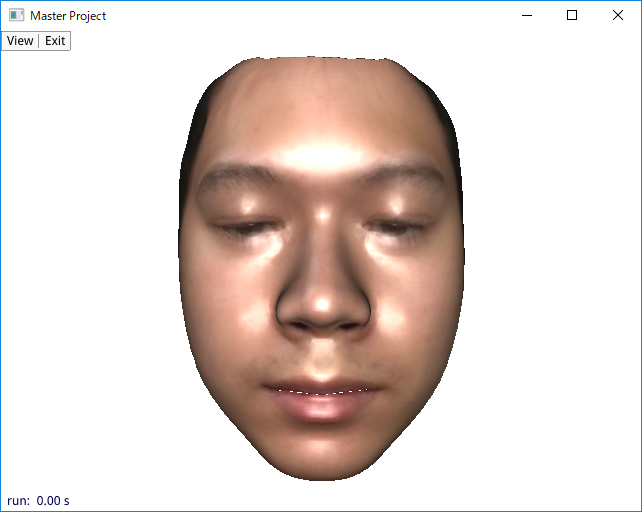}
					\hspace{3.0cm} [1] Front view
				\end{center}
			\end{minipage}
			\begin{minipage}{0.5\hsize}
				\begin{center}
					\includegraphics[clip, width=3.8cm]{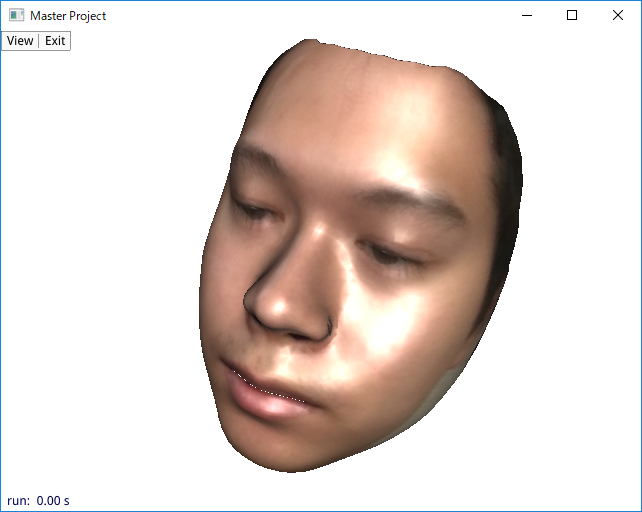}
					\hspace{3.0cm} [2] Side view
				\end{center}
			\end{minipage}
		\end{tabular}
		\caption{Interpolated 3D Face Model with morphed image.}
		\label{sixteen}
	\end{center}
\end{figure}

Given that the meshes respective to the input images have the same connectivity and number of vertices,
a target interpolated 3D model can be simply obtained by linear interpolation of the vertices of the two input meshes.
Fig.\ref{sixteen} illustrates one interpolated 3D face model with its corresponding morphed texture image computed in the  previous step.

\section{Results}

\begin{figure}[H]
	\begin{center}
		\includegraphics[scale = 0.14]{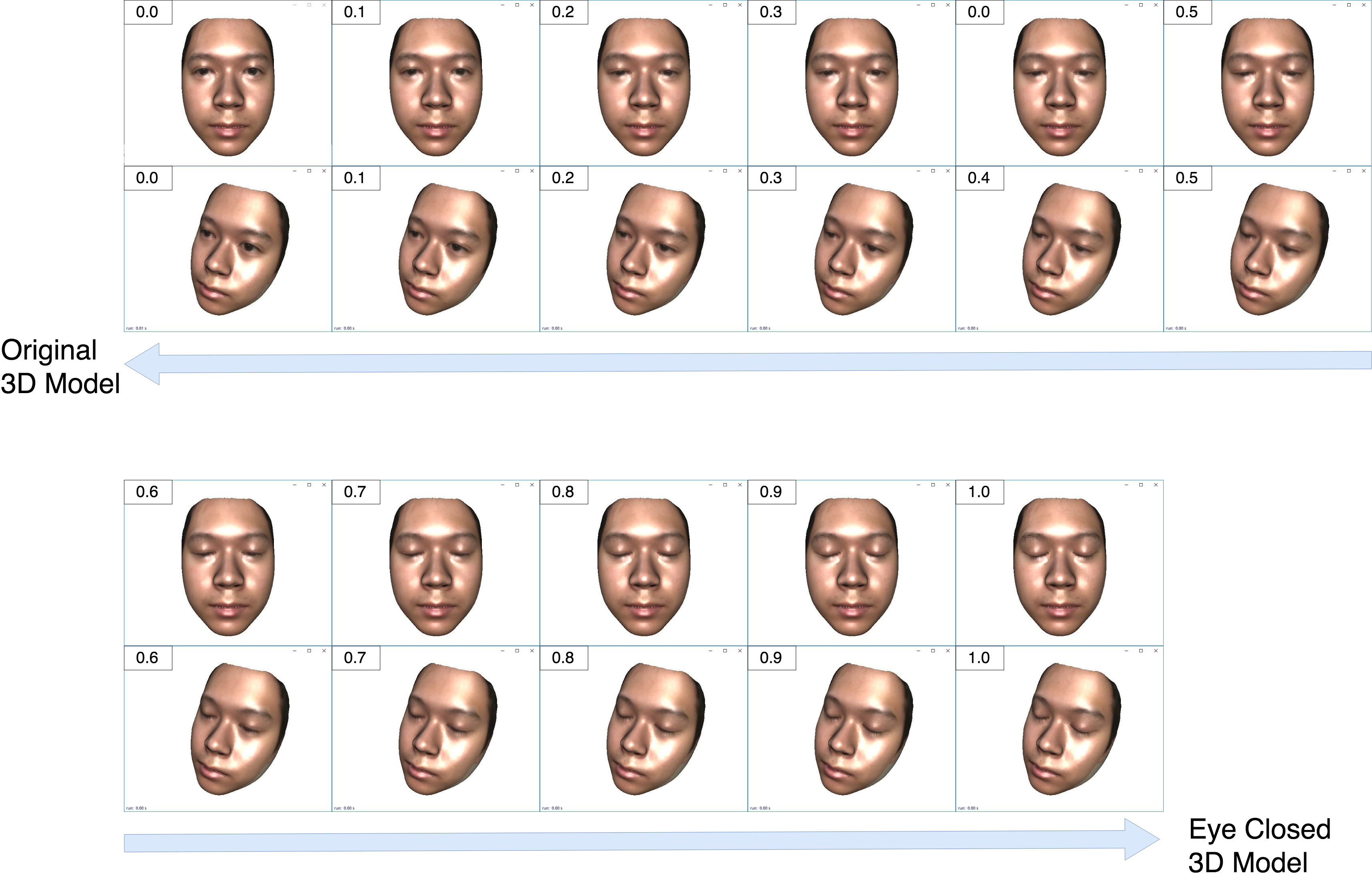}
		\caption{Pair 1. Upper Left: original 3D model. Lower Right: eye closed 3D model.}
		\label{seventeen}
	\end{center}
\end{figure}

\begin{figure}[H]
	\begin{center}
		\includegraphics[scale = 0.14]{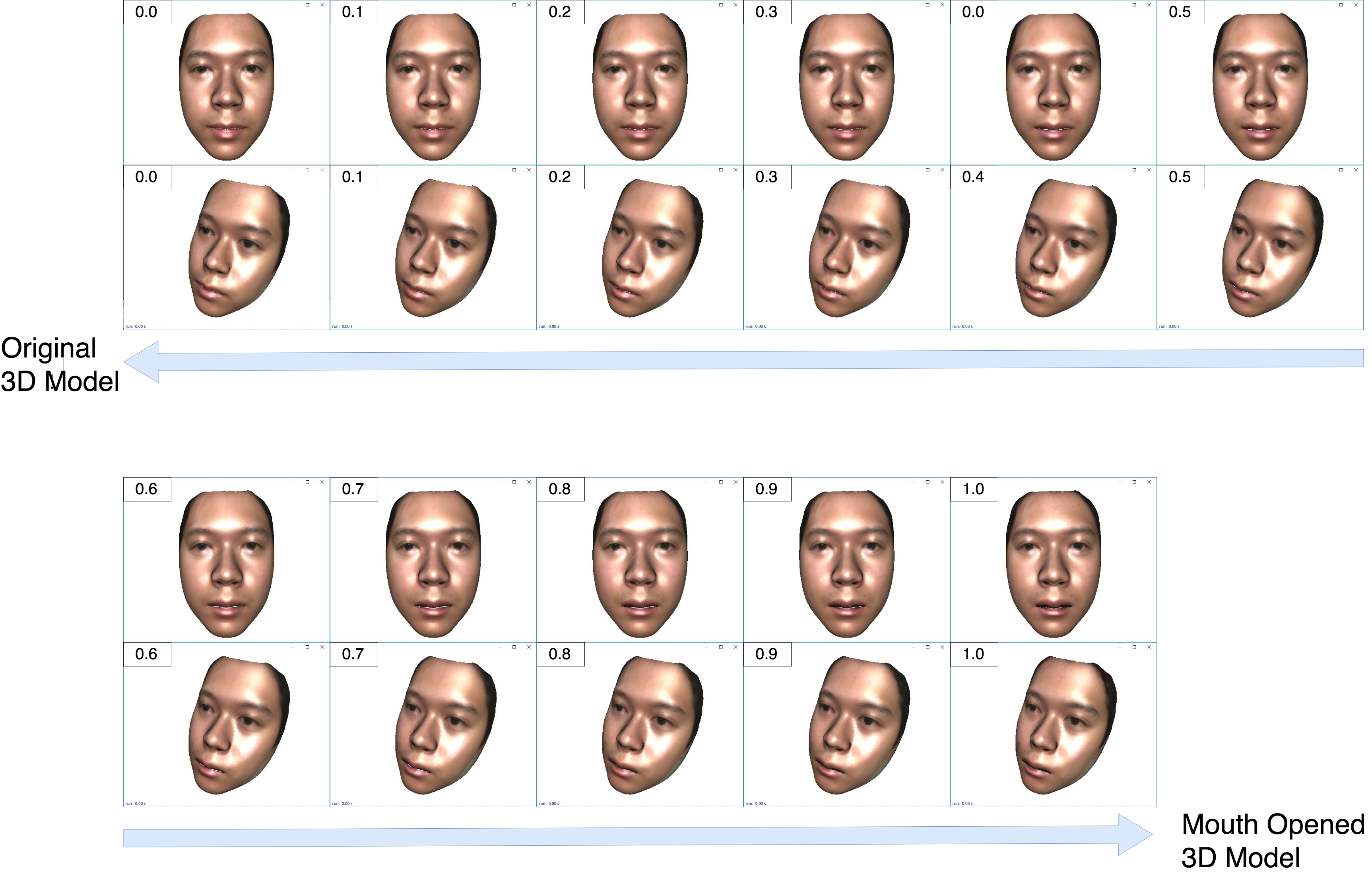}
		\caption{Pair 2. Upper Left: original 3D model. Lower Right: mouth opened 3D model.}
		\label{eighteen}
	\end{center}
\end{figure}

\begin{figure}[H]
	\begin{center}
		\includegraphics[scale = 0.14]{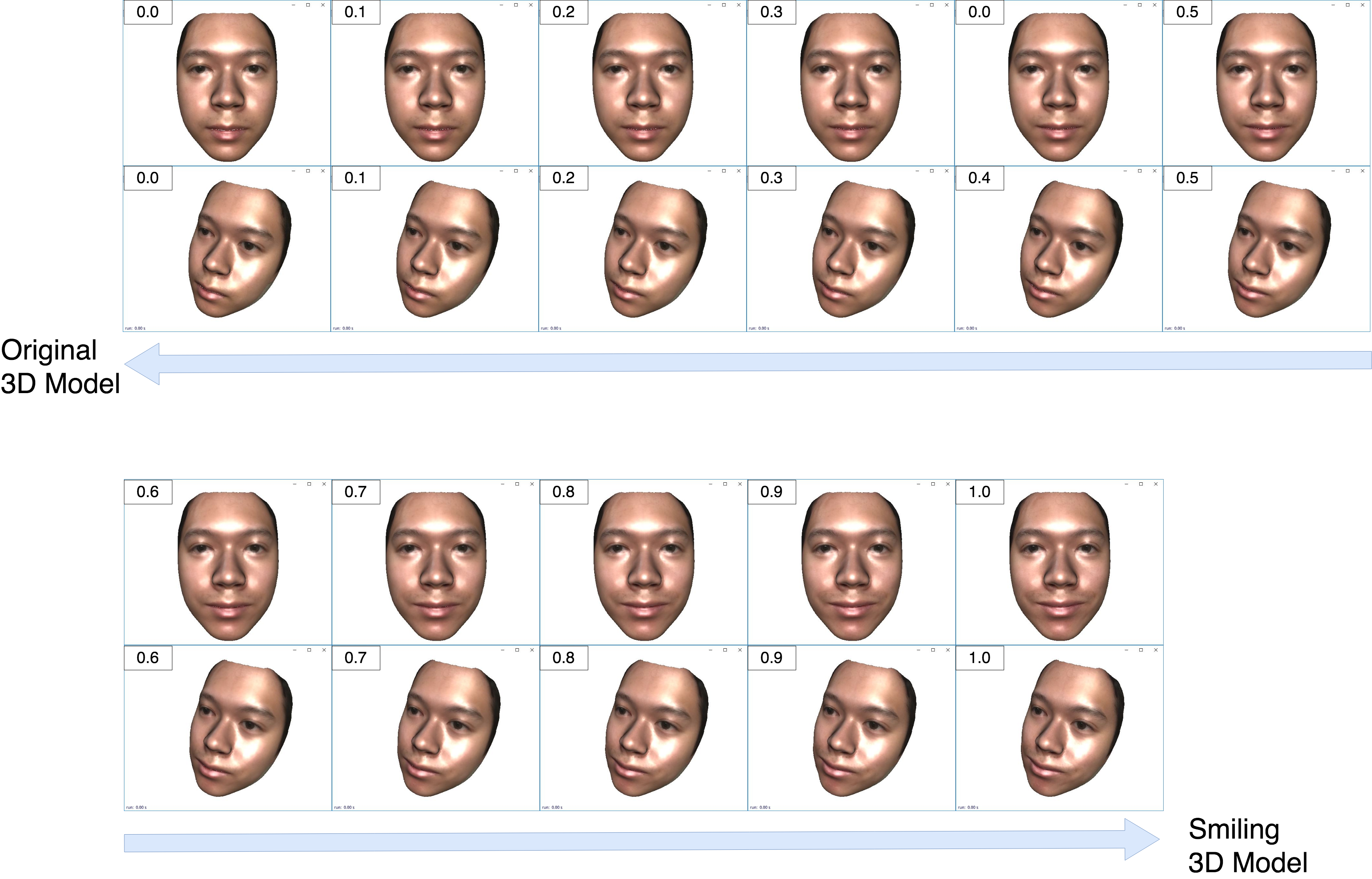}
		\caption{Pair 3. Upper Left: original 3D model. Lower Right: smiling 3D model.}
		\label{nineteen}
	\end{center}
\end{figure}

\begin{figure}[H]
	\begin{center}
		\includegraphics[scale = 0.14]{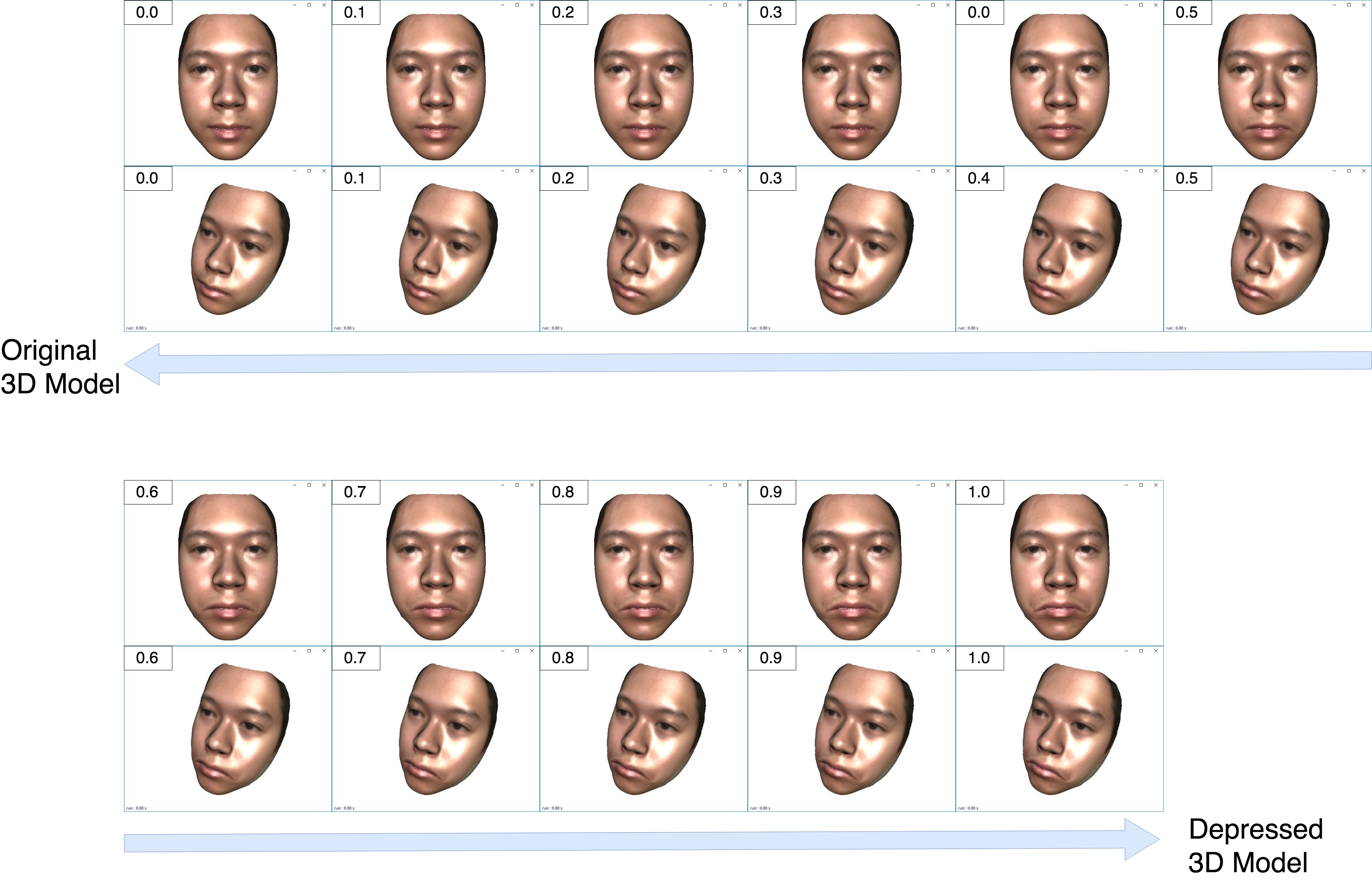}
		\caption{Pair 3. Upper Left: original 3D model. Lower Right: depressed 3D model.}
		\label{twenty}
	\end{center}
\end{figure}

Figures \ref{seventeen} to \ref{twenty} at the end of this paper illustrate the interpolated 3D face models obtained for all 4 pairs that were considered. 
The presented in-between models were obtained with interpolation parameter $t$ starting from 0 to 1, by 0.1 increments.


The morphed 3D models for all pairs look natural both with respect to the mesh and texture image deformations.
Table 1 presents performance measurements showing how long it takes to compute one interpolated 3D model frame at different resolutions with a growing number of vertices.
A 3D face model with 29587 vertices is still processed fast enough to produce smooth animations.

\begin{table}[H]
	\begin{center}
		\includegraphics[scale = 0.8]{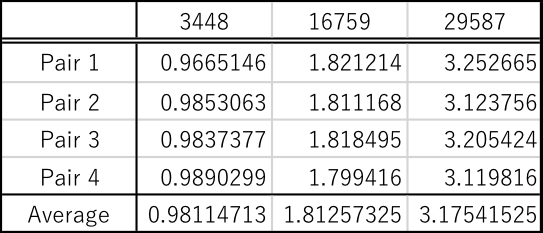}
		\caption{Time for computed one interpolated mesh (in milliseconds).}
		\label{}
	\end{center}
\end{table}

\section{Discussion}

The used Surrey Facial Model~\cite{eos} allowed us 
to construct 3D face models that were natural; however, it did not produce good results for all types of faces.
Problems were encountered especially for people with rounded faces.
The Surrey Facial Model was constructed by analyzing 169 scan data sets and nearly 60\% of the scanned data were from Caucasian people.
This might be the reason why it was difficult for the Surrey Facial Model to achieve a precise face scaling for all types of faces.

While the obtained results were smooth and of good quality, inorder to achieve a complete result additional facial objects have also to be considered. 
For example, models for eyeballs, teeth, tongue, eyebrows, eyelashes and hair are important for achieving complete animated faces.

Improved rendering techniques are also important to be considered. 
The presented results were rendered only using a standard Phong illumination model with a single frontal light source.
Special skin illumination characteristics are important to be included with dedicated shaders in order to replicate skin properties and achieve improved illumination results.


\section{Conclusion}

The described approach presents a simple pipeline to achieve fast and natural 3D facial animation between given expressions.
The approach can be replicated with the use of open source tools and animations can be obtained just from input face photos of different expressions.

\end{multicols*}

\end{document}